\documentclass[article]{aa} 
\pdfoutput=1

\usepackage{graphicx}
\usepackage{txfonts}
\usepackage[]{natbib}
\usepackage{multirow}
\usepackage{amsmath}
\usepackage{soul}
\usepackage[colorlinks=true,linkcolor=blue,citecolor=blue,draft=false]{hyperref}
\def\deg{$^{\circ}$}

\usepackage{tablefootnote} 
\usepackage{xcolor}
\usepackage{siunitx}

\newcommand{\mum}{\SI{}{\micro\meter}} 

\begin{document} 

\title{Morphology of the gas-rich debris disk around HD\,121617 with SPHERE observations in polarized light}
\author{
Cl\'ement~Perrot\inst{\ref{lesia},\ref{ifa},\ref{npf}}, 
Johan~Olofsson\inst{\ref{mpia},\ref{ifa},\ref{npf}},
Quentin~Kral\inst{\ref{lesia}},
Philippe~Thébault\inst{\ref{lesia}},
Mat\'ias~Montesinos\inst{\ref{vina},\ref{npf}},
Grant~Kennedy\inst{\ref{warwick},\ref{warwick2}},
Amelia~Bayo\inst{\ref{esog},\ref{ifa},\ref{npf}},
Daniela Iglesias\inst{\ref{leeds}},
Rob~van~Holstein\inst{\ref{esos}} \and
Christophe~Pinte\inst{\ref{monash},\ref{ipag}}
}
\institute{LESIA, Observatoire de Paris, Université PSL, CNRS, Sorbonne Université, Université Paris Cité, 5 place Jules Janssen, 92195 Meudon, France\label{lesia}
\and Instituto de F\'isica y Astronom\'ia, Facultad de Ciencias, Universidad de Valpara\'iso, Av. Gran Breta\~na 1111, Valpara\'iso, Chile\label{ifa}
\and N\'ucleo Milenio Formaci\'on Planetaria - NPF, Universidad de Valpara\'iso, Av. Gran Breta\~na 1111, Valpara\'iso, Chile\label{npf}
\and Max Planck Institute for Astronomy, K\"onigstuhl 17, D-69117 Heidelberg, Germany\label{mpia}
\and Escuela de Ciencias, Universidad Vi\~na del Mar, Vi\~na del Mar, Chile \label{vina}
\and Department of Physics, University of Warwick, Gibbet Hill Road, Coventry, CV4 7AL, UK\label{warwick}
\and Centre for Exoplanets and Habitability, University of Warwick, Gibbet Hill Road, Coventry CV4 7AL, UK\label{warwick2}
\and European Southern Observatory, Karl-Schwarzschild-Strasse 2, 85748 Garching bei München, Germany\label{esog}
\and School of Physics and Astronomy, Sir William Henry Bragg Building, University of Leeds, Leeds LS2 9JT, UK \label{leeds}
\and European Southern Observatory, Alonso de Cordova 3107, Vitacura, Casilla 19001, Santiago, 8320000, Chile\label{esos}
\and School of Physics and Astronomy, Monash University, Vic 3800, Australia\label{monash}
\and Univ. Grenoble Alpes, CNRS, IPAG, F-38000 Grenoble, France\label{ipag}
} 
\date{Accepted 2023/02/06}
\offprints{clement.perrot'at'obspm.fr}

\titlerunning{HD121617 in polarized light with SPHERE}

\authorrunning{Perrot et al.}

\abstract
	{Debris disks are the signposts of collisionally eroding planetesimal circumstellar belts, whose study can put important constraints on the structure of extrasolar planetary systems. The best constraints on the morphology of disks are often obtained from spatially resolved observations in scattered light. Here, we investigate the young ($\sim$16\,Myr) bright gas-rich debris disk around HD\,121617.}
	{We use new scattered-light observations with VLT/SPHERE to characterize the morphology and the dust properties of this disk. From these properties we can then derive constraints on the physical and dynamical environment of this system, for which significant amounts of gas have been detected.}
	{The disk morphology is constrained by linear-polarimetric observations in the \textit{J} band. Based on our modeling results and archival photometry, we also model the SED to put constraints on the total dust mass and the dust size distribution. We explore different scenarios that could explain these new constraints.}
	{We present the first resolved image in scattered light of the debris disk HD\,121617. We fit the morphology of the disk, finding a semi-major axis of $78.3\pm0.2$ au, an inclination of $43.1\pm0.2^{\circ}$ and a position angle of the major axis with respect to north, of $239.8\pm0.3^{\circ}$, compatible with the previous continuum and CO detection with ALMA. Our analysis shows that the disk has a very sharp inner edge, possibly sculpted by a yet-undetected planet or gas drag. While less sharp, its outer edge is steeper than expected for unperturbed disks, which could also be due to a planet or gas drag, but future observations probing the system farther from the main belt would help explore this further. The SED analysis leads to a dust mass of $0.21\pm0.02\,M_{\oplus}$ and a minimum grain size of $0.87\pm0.12\,\mum$, smaller than the blowout size by radiation pressure, which is not unexpected for very bright collisionally active disks.}
    {}
 
\maketitle

\section{Introduction}
\label{section1}

Debris disks are circumstellar disks orbiting $\gtrsim$ 10 Myr stars. They are detected around all types of stars of all ages \citep[e.g. ][]{2007ApJ...663..365W,2018MNRAS.475.3046S} and are found around up to $\sim$ 75\% of the stars in the youngest moving groups \citep[e.g. the F stars in the $\beta$ Pic moving group,][]{2021MNRAS.502.5390P}. Debris disks are observed via $\leq 1$mm dust produced by destructive collisions between solid bodies up to $\sim$1-100s of km in size \citep{2021MNRAS.500..718K}. Most of the disk's mass is contained in the largest of these solid bodies which constitute planetesimal belts similar to the Kuiper belt objects (KBOs) in our Solar System. Those large bodies cannot be observed in extrasolar systems. However, the location of these belts can be probed in the mm (e.g. with ALMA), which targets the thermal emission of large mm-sized grains that are unaffected by stellar radiation pressure and thus have a dynamics similar to their larger parent bodies \citep[e.g.][]{2019AJ....157..135M}. In contrast, scattered-light observations \citep[e.g. at near-infrared wavelengths with SPHERE,][]{2019A&A...626A..95P} probe much smaller micron-sized dust, whose orbits are strongly affected by radiation pressure and whose location might strongly depart from that of the planetesimal belt \citep{thebault2014}.

Detecting and spatially resolving debris disks in scattered light using the extreme adaptive optics high-contrast imagers \citep[SPHERE, GPI, SCExAO][]{2019A&A...631A.155B,2014PNAS..11112661M,2015PASP..127..890J} or space telescopes \citep[e.g., HST][]{2003SPIE.4854...81F} has now become common \citep[e.g.][]{2020AJ....160...24E,2014AJ....148...59S,2017A&A...601A...7F,2016A&A...586L...8L,2016A&A...590L...7P}. Linearly polarized light can also be observed with the Spectro Polarimetric High contrast Exoplanet REsearch (SPHERE) instrument, which provides an alternative and sensitive method to detect debris disks surrounding bright stars emitting unpolarized light \citep[e.g.][and others...]{2017A&A...607A..90E,2016A&A...591A.108O,2020AJ....160...79A}. 
These polarimetric observations reach a contrast close to the photon-noise limit \citep[][appendix E]{2021A&A...647A..21V} and the extracted disk parameters do not suffer from self-subtraction effects \citep{2012A&A...545A.111M} as are seen with angular differential imaging \citep[ADI,][]{2006ApJ...641..556M} in total intensity.
Polarized light provides additional information on the optical properties of dust compared to the total-intensity signal \citep[e.g.][]{2019A&A...626A..54M,2021A&A...653A..79S,2021ApJ...915...58C}. Thanks to the very high contrast and spatial resolution of SPHERE, remarkably sharp and deep images of debris disks can be obtained \citep[e.g.][]{2018A&A...614A..52B,2018A&A...617A.109O}, which provide information on the spatial distribution of dust that can be used to derive important information about the global structure of the planetary system \citep[e.g.][]{2016ApJ...827..125L}. As an example, the detection of a sharp inner edge to the disk could be interpreted as the signature of a planet located close to the edge that has cleared all dust from its chaotic zone \citep[e.g.][]{1980AJ.....85.1122W,2006MNRAS.373.1245Q,lagrange2012}. 

Another potential mechanism that could shape the radial structure of a debris ring, with possible consequences on its inner and outer edges, is the drag due to gas, as it will selectively make small dust grains drift inward or outward depending on their sizes \citep{2001ApJ...557..990T,2022MNRAS.513..713O}. These considerations are important because gas is now being detected in debris disks \citep[e.g.][]{2012ApJ...758...77Z,2018ARA&A..56..541H}. In fact, mainly thanks to the Atacama Large Millimeter and Submillimeter Array (ALMA), it has recently been shown that the presence of gas in a young bright debris disk is the norm rather than the exception \citep{2017ApJ...849..123M}. The observed gas (mainly CO, carbon and oxygen) is thought to be released from volatiles contained initially in icy form in the planetesimals of these belts, providing access to the composition of the volatile phase of these KBO-like bodies \citep[e.g.][]{2012ApJ...758...77Z, 2016MNRAS.461..845K}. In this scenario, both dust and gas would be of secondary origin. However, for the most massive gas disks, and only those, it is still possible that the observed gas is a relic of the protoplanetary disk phase that takes longer than expected to dissipate \citep[e.g.][]{2013ApJ...776...77K,2021ApJ...915...90N}. Such a primordial origin is however not necessary because the presence of large amounts of CO can also be explained by gas released by planetesimals creating a layer of neutral carbon gas shielding CO from photodissociating \citep{2019MNRAS.489.3670K}. Some observational evidence seems to make the primordial origin less convincing than the secondary hypothesis \citep{2017ApJ...839...86H, 2022MNRAS.510.1148S}. In systems with a significant amount of gas, interactions between gas and dust may become important and affect the dust size distribution at the smallest sizes, which may leave observable imprints \citep[e.g.][]{2019A&A...630A..85B, 2019ApJ...884..108M, 2022MNRAS.513..713O}.

In this paper, we study the debris disk around the young A1V star HD 121617, which is also surrounded by a massive gas disk \citep{2017ApJ...849..123M}. More precisely, this new study will present the first resolved scattered light observations of the debris disk around HD 121617, obtained at \textit{J} band, in polarization with SPHERE. The debris disk has been known for about 25 years and has now been observed at a range of wavelengths from optical to mm \citep{1998ApJ...497..330M,2012yCat.2311....0C,2017ApJ...849..123M}. Its fractional luminosity is close to $5\times 10^{-3}$ \citep{2011ApJ...740L...7M}, which places it among the brightest disks observed to date. More information on the star and its disk is provided in section \ref{section2}. In section \ref{section3}, we present the new SPHERE observations of HD 121617. In section \ref{section4}, we present the morphological analysis of the disk based on our resolved observations. In section \ref{section5}, we present the spectral energy distribution (SED) analysis of the system, which allows us to recover information on the dust size distribution and total dust mass. Finally, in section \ref{section6}, we discuss our results before concluding in section \ref{section7}.

\section{HD 121617}
\label{section2}

\subsection{Stellar parameters}
HD\,121617 is an A1V \citep{1978mcts.book.....H} main sequence star \citep[]{2018ApJ...859...72M}, member of the Upper Centaurus Lupus (UCL) association \citep[]{2000MNRAS.313...43H,2018ApJ...860...43G}, with an age estimated to be $16$\,$\pm$\,2\,Myr, based on the age of the UCL association \citep{2016MNRAS.461..794P}. The distance of the star is $117.9$\,$\pm$\,0.5\,pc \citep[DR3,][]{2016A&A...595A...1G,2022arXiv220605536G}.
The most recent estimations of the stellar parameters are reported in different studies, for example in \citet{2018A&A...614A...3R} where the authors estimated the effective temperature $T_\mathrm{eff}=9285\,K$ and surface gravity of $\log\,g=4.45$. In \citet{2016ApJS..225...15C}, the effective temperature is estimated at $T_{eff}=8710\,K$ and a stellar radius at $R_{\star}=1.63\,R_{\odot}$, while \citet{2018ApJ...859...72M} estimated the stellar luminosity at $L_{\star}=17.3\,L_{\odot}$ and the stellar mass at $M_{\star}=1.9\,M_{\odot}$.
The galactic extinction is estimated by \textit{Gaia} DR3 to $A_{0}=0.1331^{+0.0012}_{-0.0018}$\,mag \citep{2016A&A...595A...1G,2022arXiv220605536G} and the ratio between the extinction and the reddening to $R_{V}=3.25$ \citep{2018MNRAS.475.1121G}.

\subsection{Debris disk in the infrared}
The presence of a disk around HD\,121617 was first reported in \citet{1998ApJ...497..330M} as an infrared excess at $12$, $25$, $60$ and $100\,\mum$ in the Infrared Astronomical Satellite (IRAS) Faint Source Survey Catalog \citep{1989ifss.book.....M}.
The first estimation of the temperature and radius of the disk was made by \citet{2013A&A...550A..45F} using AKARI/IRC observations at $18\,\mum$ \citep{2010A&A...514A...1I}, on top of the previous Infrared Astronomical Satellite (IRAS) data as well as the Wide-field Infrared Survey Explorer (WISE) survey \citep{2012yCat.2311....0C}. 
\citet{2011ApJ...740L...7M} reported a fractional luminosity of the disk $f_\mathrm{disk} = L_\mathrm{disk}/L_\star =4.8\times10^{-3}$, based on the SED.

\subsection{Gas and dust detection with ALMA}
The first millimeter detection was reported in \citet{2017ApJ...849..123M} with ALMA at $1.3$\,mm. They spatially resolved the disk in the continuum and detected spectrally resolved emission lines of several CO isotopologs, showing the presence of gas in the debris disk. The dust mass is estimated to be $1.4$\,$\times$\,$10^{-1}$\,M$_{\oplus}$ from the $1.3$\,mm observations. They also constrained the morphology of the ring from the continuum observations (but not for the gas), with an inclination of $37$\,$\pm$\,$13^{\circ}$, a position angle of the major axis of $43$\,$\pm$\,$19^{\circ}$, a diameter of $152$\,$\pm$\,$15$\,au, and a radial thickness of $52$\,$\pm$\,$17$\,au, corresponding to the full width at half maximum (FWHM) of a the 2D-Gaussian used to model the ring \citep{2017ApJ...849..123M}. The ring size and thickness are corrected, according to the new distance of the star from Gaia DR3. As for the gas observations, using the standard isotopolog ratios of the local interstellar medium, they estimated a total $^{12}$CO mass of $1.8$\,$\times$\,$10^{-2}$\,M$_{\oplus}$, which makes it part of the most massive gas disks detected so far in debris disk systems, with a gas-to-dust ratio of $\sim$0.13.


\section{Observations}
\label{section3}

\subsection{SPHERE data}
\label{section3.1}

\begin{figure*}[ht] 
\includegraphics[width=0.99\textwidth,clip]{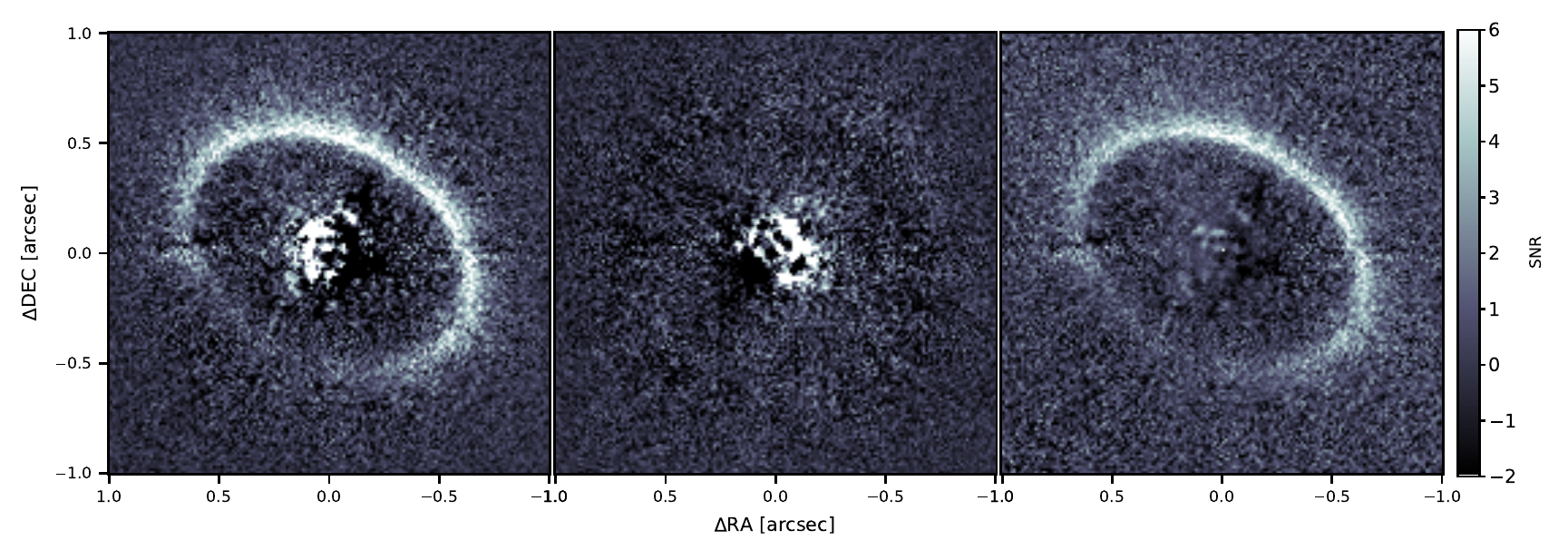}
\normalsize
\caption{Left to right: $Q_\phi$ image, $U_\phi$ image, and S/N map of the $J$ band observations of HD\,121617. North is up, east is left.}
\label{observation}
\end{figure*}

HD\,121617 was observed with VLT/SPHERE \citep{2019A&A...631A.155B} on the 28$^\mathrm{th}$ of April 2018 and the 20$^\mathrm{th}$ of May 2018 with the same configuration\footnote{ESO program ID: 0101.C-0420(A), PI: Johan Olofsson.}.
Both epochs used the dual-beam polarimetric imaging \citep[DPI,][]{2020A&A...633A..63D, 2020A&A...633A..64V} mode of the Infra-Red Dual-band Imager and Spectrograph \citep[IRDIS, ][]{2008SPIE.7014E..3LD}. 
The observations were done with the broadband J filter (BB\_J, $\lambda_c$\,=\,1.245\,$\mum$, $\Delta\lambda$\,=\,240\,nm, \citealp{2020A&A...633A..63D}), the N\_ALC\_YJH\_S coronagraph mask, with an inner working angle of 185\,mas \citep{2008SPIE.7015E..1BB} and the pixel scale for this configuration is 12.26\,mas per pixel \citep{2016SPIE.9908E..34M}. The observations were performed in field tracking mode and an offset angle was applied in the derotator to avoid the loss of polarisation as described in \citet{2020A&A...633A..63D}.

The DPI mode of IRDIS allows to construct the images of Stokes $Q$ and $U$. For this, the light is split into two parallel beams which pass two linear polarizers with orthogonal transmissions and are imaged simultaneously on the same detector, in the so-called Left and Right area. The subtraction of those images (Left minus Right) gives four different $Q^{\pm}$ and $U^{\pm}$ images, depending on the angle orientation of the half-wave plate (HWP). When the HWP switch angle is 0\deg, the $Q^+$ image is formed. HWP angles of 22.5\deg, 45\deg and 65.5\deg form respectively the images $U^+$, $Q^-$ and $U^-$. Finally, the Stokes $Q$ and $U$ images are constructed following:
\begin{equation}
    X = \frac{1}{2}(X^+ - X^-)
\end{equation}
where $X$ equals $U$ to reconstruct the Stokes $U$ vector and $X$ equals $Q$ to reconstruct the Stokes $Q$ vector.
Thereby, a cycle of 4 HWP switch angles is necessary to obtain the Stokes $U$ and $Q$ vectors. More detailed explanations can be found in \citet{2020A&A...633A..63D} and \citet{2020A&A...633A..64V}.
A total of $24$ HWP cycles were obtained (96 frames) for each epoch. The exposure time was set to $32$ seconds (DIT) for a trade-off between a maximum of time integration and avoiding saturation around the coronagraph, for a total integration time of $3072$ seconds (51.2 minutes) for each epoch.
The observation sequence was performed following this order. First we acquired non-coronagraphic and non-saturated frames of the star (shifted out of the coronagraph by few hundreds mas, for the flux calibration. Then, the star is moved back behind the coronagraph to perform frame centering with the ``waffle'' mode, a sinusoidal pattern put on the deformable mirror in order to create 4 symmetric spots. The intersection of those 4 spots gives with a high accuracy the position of the star behind the coronagraph. The science acquisition is then obtained, with the polarimetric cycles described previously. At the end of the science sequence, another centering and flux calibration are performed in order to check the stability of the star centering and relative flux. Finally, a series of background calibration are performed by pointing the telescope away from the star. Both observations were taken with good atmospheric conditions with a seeing between $0.45$\arcsec and $0.75$\arcsec and a coherence time between $3$\,ms and $8.5$\,ms.


Data reduction was performed using the IRDIS Data reduction for Accurate Polarimetr (\texttt{IRDAP}\footnote{\url{https://irdap.readthedocs.io}}) pipeline \citep{2020A&A...633A..64V}.
The pipeline includes background and flat-field calibrations, star centering, correction for instrumental polarization and polarization crosstalk and the creation of the Stokes $Q$ and $U$ images. Then, \texttt{IRDAP} constructs the $Q_{\phi}$ and $U_{\phi}$ images \citep{2006A&A...452..657S}, which are used for further interpretation, following the definitions of \citet{2020A&A...633A..63D}:
\begin{equation}
    \begin{split}
        Q_{\phi} = -Q \cos{2\Phi} - U \sin{2\Phi}\\
        U_{\phi} = +Q \sin{2\Phi} - U \cos{2\Phi}
    \end{split}
\label{eq_q_u_phi}
\end{equation}
where 
$\Phi$ is the position angle of the location of interest with respect to the stellar location.  
Figure \ref{observation} shows the final $Q_{\phi}$ and $U_{\phi}$ images obtained (the mean of the two epochs). 
The $Q_{\phi}$ image contains the polarimetric signal of the disk, revealing a bright and narrow ring. The ring shows a flux asymmetry between the North-West and the South-East sides. The $U_\phi$ image does not show any structured signal, expected for an optically thin disk \citep{2015A&A...582L...7C}, and can therefore be used as a proxy for uncertainties in the modeling of the observations.
Figure \ref{observation} (right) shows the derived signal-to-noise (S/N) map of $Q_{\phi}$.
Several artifacts remain in the $Q_{\phi}$ and $U_{\phi}$ reduced images. In the East part of both images ($\Delta$RA\,=\,$0.5$\arcsec, $\Delta$DEC\,=\,$0$\arcsec) an horizontal artifact can be seen. This artifact, due to the SPHERE deformable mirror \citep[fitting error,][]{2019Msngr.176...25C}, is also present on the opposite side of the image, but less visible. Another artifact along the diagonal (from the top-right to the bottom-left) is present and is most likely due to the diffraction pattern of the VLT's spiders, which were not aligned with the Lyot stop in field-tracking mode \citep[low-order residuals,][]{2019Msngr.176...25C}. Fortunately, the impact of those artifact is small due to the brightness of the disk and their locations.

\subsection{Photometry}
\label{section3.2}

Using the VO SED Analyzer (\texttt{VOSA}\footnote{\url{http://svo2.cab.inta-csic.es/theory/vosa/}}) \citep{2008A&A...492..277B} we gathered the photometric observations from different missions and instruments to build the SED of HD\,121617.
To determine the stellar parameters (section \ref{section2}) we used the photometric measurements from TYCHO \citep[B, V filters,][]{2000A&A...355L..27H}, Gaia \citep[Gbp, G and Grp filters,][]{2018yCat.1345....0G}, 2MASS \citep[J, H and Ks filters,][]{2003tmc..book.....C} and WISE \citep[W1 and W2 filters,][]{2010AJ....140.1868W}. While WISE \citep[W3 and W4 filters,][]{2010AJ....140.1868W}, Herschel/PACS (, $100\,\mum$ and $160\,\mum$, this work) and the 1.3\,mm ALMA observations \citep{2017ApJ...849..123M} are used to fit the infrared excess (section \ref{section5}). 
We corrected the photometry to the extinction estimated by Gaia ($A_V=0.134$ mag, converted from $A_0$), assuming $R_V=3.25$ from \citet{2018MNRAS.475.1121G}. We converted $A_V$ to $A_{\lambda}$ for each filters using the extinction law from \citet[][Eq. 1, 2 and 3]{1989ApJ...345..245C} for wavelengths below $3.3\,\mum$ and the extinction law from \citet[][$A_{\lambda}=(\frac{0.55}{\lambda [\mum]})^{1.6} \times A_V$]{2003ApJ...584..853P} for wavelengths larger than $3.3 \mum$.
The photometric points, corrected and not corrected for extinction, are presented in Table\,\ref{sedvalue}.

\begin{figure}[] 
\includegraphics[width=0.49\textwidth,clip]{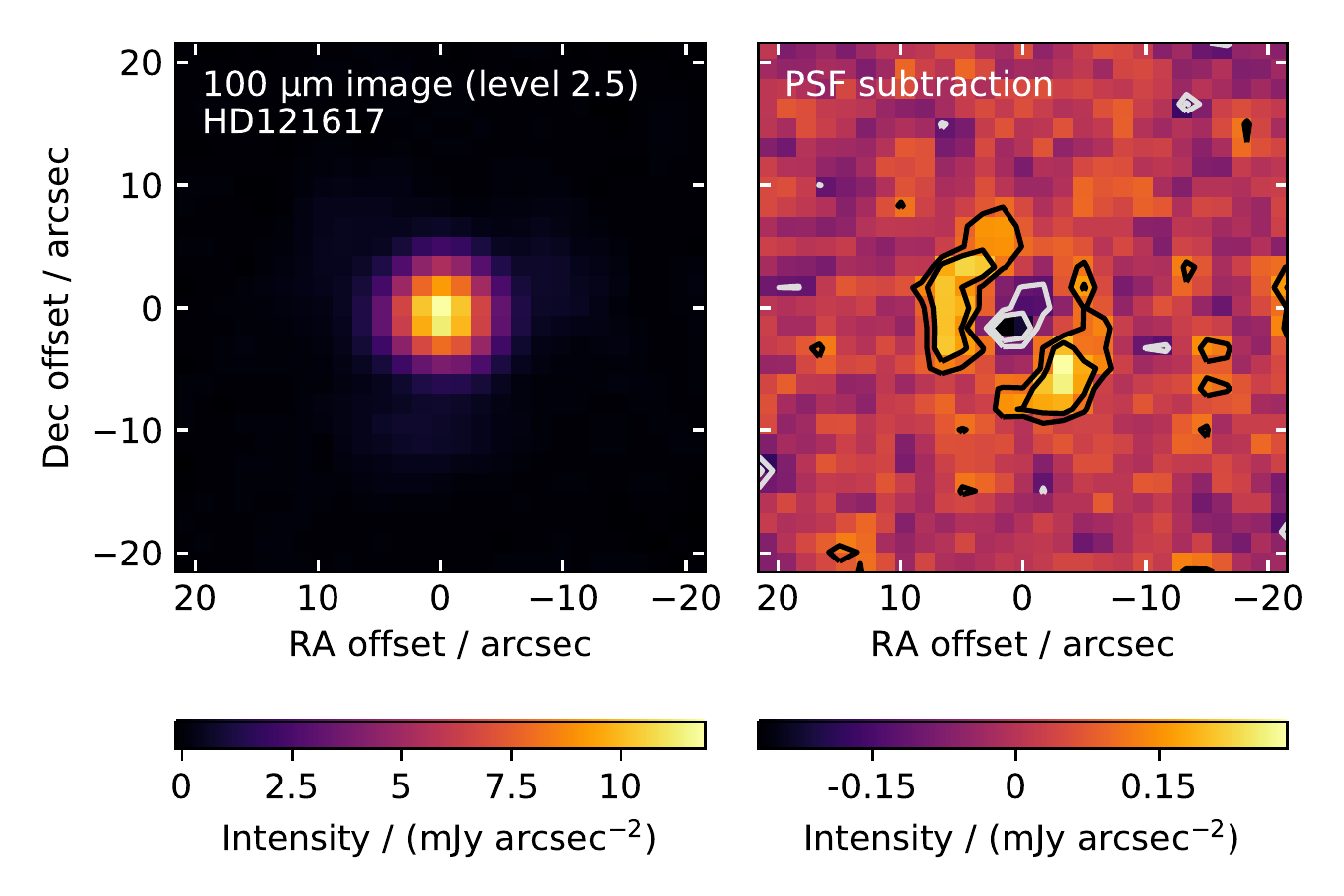}
\normalsize
\caption{Left: Herschel/PACS observation of HD\,121617 at 100\,$\mum$. Right: Residual of the left image after a PSF subtraction performed with a calibration star. The continuum of the disk is marginally resolved.}
\label{pacsobs}
\end{figure}

The 100\,$\mum$ PACS image is in fact marginally resolved; Figure \ref{pacsobs} shows the data and a residual plot made by subtracting a point spread function (PSF) that was scaled to the image peak (where the PSF is an observation of the calibration star $\gamma$~Dra). Fitting the residual image with a disk model \citep[as in][]{2019MNRAS.488.3588Y} finds disk parameters that are consistent with those found in section \ref{section4}, but with significantly greater uncertainties, so we do not use this spatial information for any further analysis.

\begin{table}[ht!]
\caption{Photometric points for HD\,121617.}
\footnotesize
\centering
\begin{tabular}{ l r r r r r}
\hline 
\hline
Filter & Flux & Flux corr. & Error & $\lambda_{\rm eff}$ & Ref.\\
          - &     [mJy] & [mJy] &   [mJy] & [$\mum$] & - \\
\hline
 TYCHO $B$    &   4372.0 & 5166.1 & 64.79 & 0.428 & 1\\
 GAIA $Gbp$  &   3993.0 & 4577.1 & 37.59 & 0.504 & 2\\
 TYCHO $V$    &   4499.0 & 5111.1 & 48.14 & 0.534 & 1\\
 GAIA $G$    &   3957.0 & 4440.2 & 36.39 & 0.586 & 2\\
 GAIA $Grp$  &   3236.0 & 3506.8 & 30.53 & 0.769 & 2\\
 2MASS $J$    &   2067.0 & 2143.2 & 58.18 & 1.235 & 3\\
 2MASS $H$    &   1400.0 & 1431.8 & 60.48 & 1.662 & 3\\
 2MASS $Ks$   &    883.0 & 896.1 & 15.97 & 2.159 & 3\\
 WISE $W1$    &    409.0 & 411.8 & 14.73 & 3.35 & 4\\
 WISE $W2$    &    240.0 & 241.0 &  7.56 & 4.60 & 4\\
 WISE $W3$    &     72.2 & 72.3 & 3.14 & 11.56 & 4\\
 WISE $W4$    &    567.0 & 567.2 & 31.28 & 22.09 & 4\\
 PACS 100   &   961.8 & 961.8 & 24.25 & 97.90 & 5\\
 PACS 160   &   415.4 & 415.4 & 17.17 & 153.95 & 5\\
 ALMA  &    1.86 & 1.86 & 0.29 & 1330.0 & 6\\
\hline
\hline
\end{tabular}
\flushleft
\vspace{0.1cm}
\textbf{Notes.} The effective wavelengths $\lambda_{\rm eff}$ were taken from SVO Filter Profile Service\tablefootnote{\url{http://svo2.cab.inta-csic.es/theory/fps/}} \citep{2012ivoa.rept.1015R,2020sea..confE.182R}. The Flux corr. column corresponds to the Flux corrected to the extinction described in section \ref{section3.2}. Reference: 1) \citet{2000A&A...355L..27H}, 2) \citet{2018yCat.1345....0G}, 3) \citet{2003tmc..book.....C}, 4) \citet{2010AJ....140.1868W}, 5) this work, 6) \citet{2017ApJ...849..123M}.
\normalsize
\label{sedvalue}
\end{table}


\section{Morphological analysis of the SPHERE observations}
\label{section4}

\subsection{Model and method}
\label{section4.1}

In order to constrain the disk morphology, we used the \texttt{Debris DIsks Tool}\footnote{\url{https://github.com/joolof/DDiT}} \citep[\texttt{DDiT},][]{2020A&A...640A..12O} to create synthetic models of the debris disk in polarized intensity. \texttt{DDiT} was used with an Markov chain Monte Carlo (MCMC) code based on \texttt{emcee}\footnote{\url{https://emcee.readthedocs.io/en/stable/}} \citep{2013PASP..125..306F} to determine the best values of the tested parameters.
The geometry of the disk is defined by the inclination $i$ ($0^{\circ}$ for a face-on disk and $90^{\circ}$ for an edge-on disk) and the position angle of the major axis with respect to the north $\phi$.
The dust density distribution of the disk is described in the radial $r$ and vertical $z$ directions as:
\begin{equation}
\label{dust_density}
    n(r,z) \propto \left[ \left(\frac{r}{r_0} \right)^{-2\alpha_\mathrm{in}} + \left(\frac{r}{r_0} \right)^{-2\alpha_\mathrm{out}} \right]^{-1/2} \times e^{-z^2/2h^2}
\end{equation}
with $n$ the dust grain volumetric density, $r_0$ the reference radius of the disk, $\alpha_\mathrm{in}$ and $\alpha_\mathrm{out}$ the inner and outer coefficients of the slope of the dust density distribution and $h$ the scale height of the disk. In the case of a non-eccentric circular debris disk, $r_0$ is a constant equal to the semi-major axis $a$. For an eccentric orbit, $r_0$ is defined as:
\begin{equation}
    r_0(\gamma) = \frac{a(1-e^2)}{1+e\cos{(\omega + \gamma)}}
\end{equation}
with $e$ the eccentricity, $\omega$ the argument of the pericenter and $\gamma$ the azimuthal angle of the disk at $r_0$. Therefore, our fit will allow us to test whether the disk may be slightly eccentric. 
For the polarized scattering phase function we used the Henyey-Greenstein approximation \citep{1941ApJ....93...70H}, which is parameterized with the coefficient $g$ characterizing the scattering anisotropy of the dust (defined between $-1$, for backward scattering and $1$ for forward scattering) and the Rayleigh scattering function:
\begin{equation}
    f_{HG} = \frac{1-\cos^2(\theta)}{1+\cos^2(\theta)} \frac{1}{4\pi}\frac{1-g^2}{(1+g^2-2g\cos{\theta})^{3/2}}
\end{equation}

\noindent where $\theta$ is the scattering angle \citep{2017A&A...607A..90E,2019A&A...630A.142O}.
Finally, the aspect ratio, $\psi=\arctan(h/r)$, is fixed to $0.05$\,rad \citep{2009A&A...505.1269T} due to the inclination of the disk, which is not adapted for a good estimation of $\psi$. 

Priors for the 8 free parameters of the MCMC run ($a$, $i$, $\phi$, $g$, $\alpha_{\rm in}$, $\alpha_{\rm out}$, $e$, $\omega$) are presented in Table \ref{mcmcparam}.
To compare each model to the data and determine the best value of the free parameters we proceed as follows.
For each iteration of the MCMC run, the model is first convolved by a 2D normal distribution with a standard deviation of $2$ pixels (corresponding to the full width at half maximum of the off-axis image, to avoid an additional source of noise). Then the flux of the model is scaled to best match the data and the scaling factor $S_\mathrm{scale}$ is estimated as:
\begin{equation}
    S_\mathrm{scale} = \frac{\sum\frac{I_\mathrm{data}\times I_\mathrm{model}}{\sigma^2} }{ \sum\left(\frac{I_\mathrm{model}}{\sigma}\right)^2}
\end{equation}
with $I_\mathrm{data}$ the 2-dimensional $Q_{\phi}$ image of the disk, $I_\mathrm{model}$ the 2-dimensional synthetic $Q_{\phi}$ image of the disk and $\sigma$ the noise map.
This scaling is performed since \texttt{DDiT} produces images with no absolute flux for the reason that the latter depends on the total dust mass. 
The minimization is performed in an area including the disk and excluding the central part of the image where the residual star light remains. We selected the area between 0.25\arcsec\,and 1.1\arcsec\,from the star position (respectively, 29 and 129\,au in projected separation). In that way, we do not exclude the potential extended signal of the disk in the outer part. In this area, the $\chi^2$ is computed following:
\begin{equation}
    \chi^2= \left(\frac{I_\mathrm{data}-S_\mathrm{scale}\times I_\mathrm{model}}{\sigma}\right)^2
\end{equation}
Finally, this $\chi^2$ is provided as the log likelihood to the MCMC for the optimization of the free parameters. The MCMC is run with 80 walkers with a length of 500 and a burn-in fixed at 100 steps.

\subsection{Noise map and uncertainties}

The noise map $\sigma$ is given by the $U_{\phi}$ image, which contain the same noise that the $Q_{\phi}$ image, without the astrophysical signal. To build the noise map, we compute the standard deviation per pixel in a small ring centered to the star. To take account for the correlation between pixels we add to the standard deviation an inflation term as described in \citet{2021ApJ...912..115H}. This inflation term considers the spatial correlation of the PSF through the instrumental FWHM (40\,mas or 3.3 pixels) and the radial elongation of the speckles due to the filter's bandwidth. Therefore the noise map is computed as:
\begin{equation}
\sigma(r) = \sqrt{FWHM \times r \times \frac{\Delta\lambda}{\lambda_c}} \times \sigma_{std}(r)
\end{equation}
with $r$ the angular separation, FWMH = 3.3\,pixels, $\Delta\lambda$ the filter's bandwidth, $\lambda_c$ the central wavelength of the filter and $\sigma_{std}(r)$ the radial standard deviation per pixel of the $U_{\phi}$ image. This is a conservative method which increases the uncertainties from the MCMC, which are usually under-estimated. 
In addition, \citet{2021A&A...651A..71L} gives the typical error on the star center for the SpHere INfrared survey for Exoplanets (SHINE), which is 1.5\,mas ($\sim$0.18 au).

\subsection{Results of the MCMC analysis}
\label{section4.2}

\begin{figure*}[!ht]
\centering
\includegraphics[width=0.99\textwidth,clip]{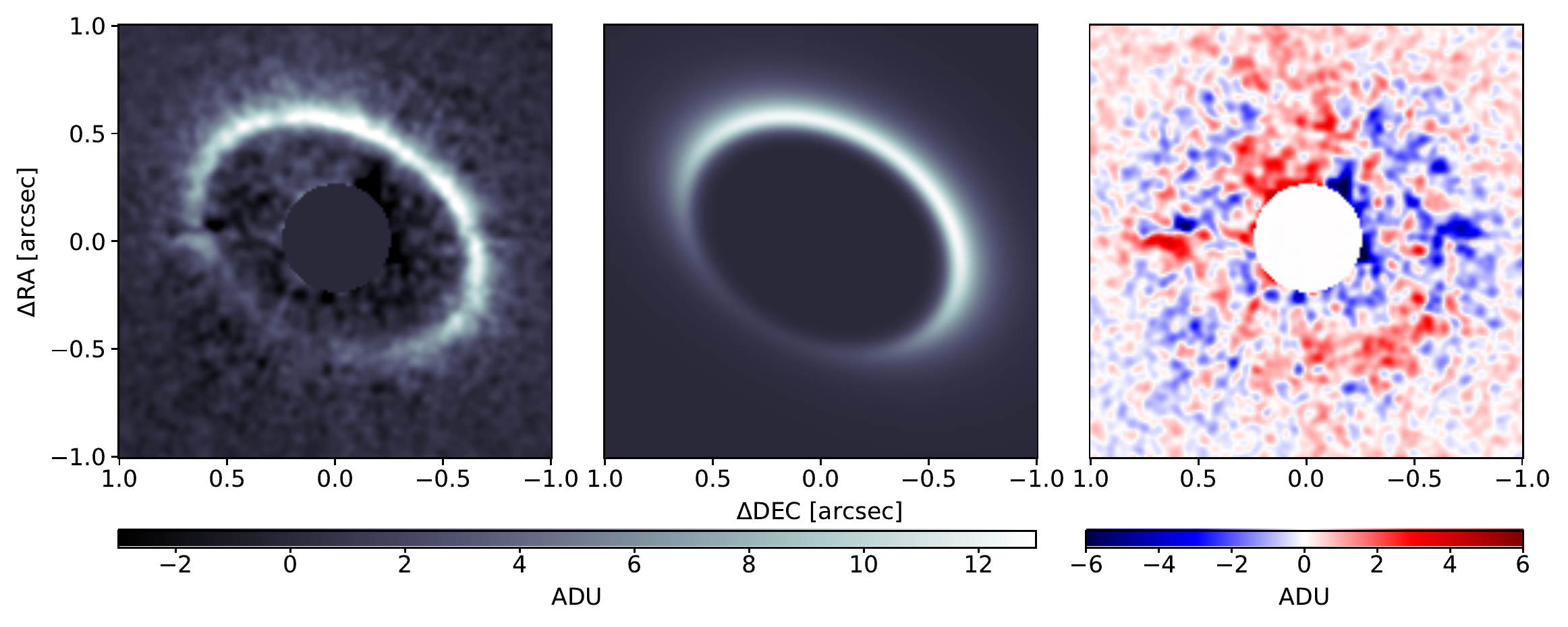}
\caption{From left to right: The $Q_{\phi}$ image, the best model and the residual image after the subtraction of the best model from the $Q_{\phi}$ image. The grey area in the middle of each image is a numerical mask to exclude the residual star light.}
\label{obsmodelresidu}
\end{figure*}

The results of the MCMC analysis are summarized in Table \ref{mcmcparam}. The best fit values are derived from the corner plots of the MCMC analysis, and their error bars are 1-$\sigma$ of the corresponding probability density function (PDF) for each given parameter (Fig. \ref{cornerplot}). 
Figure \ref{obsmodelresidu} shows the data (left), the best model (middle) and the residual image (right).
Since most of the signal from the disk is accounted for and therefore not visible on the residuals, our best-fit model can explain the observations. 

\begin{table}[ht!]
\caption{Priors and results of MCMC analysis.}
\footnotesize
\centering
\begin{tabular}{ l l r}
\hline 
\hline
Parameters & Priors   & Best-fit\\
           &         & value\\
\hline
$a$ [au]& [75 ; 95] & 78.6 $\pm$ 0.6\\
$i$ [\deg]& [30 ; 60] & 43.4 $\pm$ 0.8\\
$\phi$ [\deg]& [220 ; 260]  & 240.0 $\pm$ 0.9\\
$g$ & [0 ; 0.9]  & 0.60 $\pm $0.04\\
$\alpha_\mathrm{in}$ & [2 ; 35]  & 18.9 $\pm$ 2.3\\
$\alpha_\mathrm{out}$ & [-15 ; -2]  & -5.8 $\pm$ 0.4\\
$e$ & [0, 0.1]  & 0.03 $\pm$ 0.01\\
$\omega$ [\deg]& [60 ; 200]  & 131.0 $\pm$ 15.1\\ 
\hline
\hline
\end{tabular}
\flushleft
\vspace{0.1cm}
\textbf{Notes.} $a$: semi-major axis. $i$: inclination. $\phi$: position angle of the major axis of the disk with respect to the north. $g$: anisotropic scattering coefficient of the Henyey-Greenstien approximation for the dust. $\alpha_\mathrm{in}$ and $\alpha_\mathrm{out}$: respectively the inner and the outer slope of the power law distribution of the dust density. $e$: eccentricity. $\omega$: argument of the pericenter. The Best-fit column is derived from the corner plot (Fig. \ref{cornerplot}) of the MCMC analysis with 1-$\sigma$ error.
\normalsize
\label{mcmcparam}
\end{table}

We improve the constraints on the geometry of the disk compared to previous ALMA study \citep{2017ApJ...849..123M}. The inner edge of the disk observed with SPHERE is far enough from the star not to be affected by artifacts of residual light. Moreover, we use polarimetric observations, which do not necessitate post-processing techniques such as ADI and the results do not suffer from, for example, self-subtraction effects as can be the case for total intensity images. The SPHERE and ALMA results are compatible as we find a semi-major axis of $78.6$\,$\pm$\,$0.6$\,au (\textit{vs.} $76$\,$\pm$\,$8$\,au), an inclination of $43.4$\,$\pm$\,$0.7^{\circ}$ (\textit{vs.} $37$\,$\pm$\,$13^{\circ}$) and a position angle of the major axis of $240.0$\,$\pm$\,$0.9^{\circ}$ (\textit{vs.} $223$\,$\pm$\,$19^{\circ}$). 
However the ring observed with SPHERE is much narrower compared to that observed with ALMA.
The equivalent full width at half maximum (FWHM) of the disk in the SPHERE image is around 23\,au (0.2\arcsec), while \citet{2017ApJ...849..123M} reported a FWHM of $52\pm17$\,au (0.44\arcsec). The difference can be explained by the larger angular beam size of the ALMA observations ($\sim$ 0.5\arcsec) so that the mm-dust ring is not resolved. In contrast, in the SPHERE observations the angular resolution of 0.03\arcsec allows us to resolve the ring's width.

\begin{figure*}[] 
\includegraphics[width=0.99\textwidth,clip]{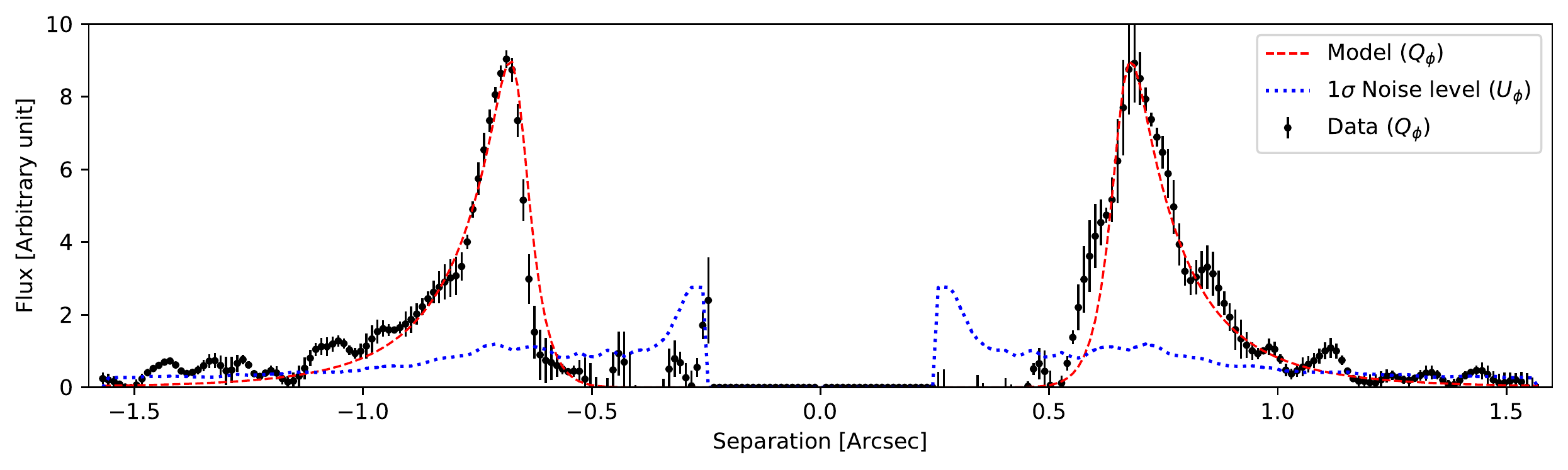}
\normalsize
\caption{$Q_{\phi}$ radial profile following the projected major axis (6 pixels width) from the South-West to the North-East of the disk with the corresponding error bars (black points). The radial profile of the best model (red dashed line) and the Noise level obtained with the azimuthal standard deviation of the $U_{\phi}$ image (blue dotted line). 
}
\label{radialprofile}
\end{figure*}

We find that the shape of the dust density distribution for the inner edge of the ring is extremely steep, with a power-law slope $\alpha_\mathrm{in}$\,=\,$18.9$\,$\pm$\,$2.3$. The outer edge of the ring is also  steeper than the expected dust density distribution for an evolved debris disk \citep{2008A&A...481..713T} with a power law of slope $\alpha_\mathrm{out}$\,=\,$-5.8$\,$\pm$\,$0.4$ instead of the typical $\alpha_\mathrm{out}$\,=\,$-1.5$. 
In Figure \ref{radialprofile}, we show the radial profiles along the major axis for the North-East and South-West directions, compared to the radial profile of the best model and the $1\sigma$ noise level from the $U_{\phi}$ image. Radial profiles are obtained by averaging a band of 6 pixels width along the major axis.

Finally, our analysis also shows that the disk is slightly eccentric, with an eccentricity of $0.03$\,$\pm$\,$0.01$. 
With a semi-major axis of $78.6$\,$\pm$\,$0.6$\,au, this leads to an apocenter at $81.0$\,$\pm$\,$1.4$\,au and a pericenter at $76.2$\,$\pm$\,$1.4$\,au from the star, with the argument of the pericenter being at $131.0$\,$\pm$\,$15.1^{\circ}$.


\section{SED modeling}
\label{section5}

The analysis made with \texttt{DDiT} was focusing on the disk morphology. In this section we use those previously constrained parameters to estimate the dust properties. This is indeed a good way to proceed because SED fitting is a degenerate problem where the radial distance of the belt and the minimum grain size are correlated with each other. But once the belt location is fixed, we can directly constrain the grain properties.

\subsection{SED with \texttt{MCFOST}}
\label{section5.1}

\texttt{MCFOST}\footnote{\url{https://github.com/cpinte/mcfost}} \citep{2006A&A...459..797P,2009A&A...498..967P} is a 3 dimensional radiative transfer code for circumstellar disks, which can produce synthetic SED, images or absorption lines for specific systems. The code was initially developed for protoplanetary disks (optically thick, with gas, in local thermal equilibrium or not), but is also suitable for optically thin debris disks.
We used \texttt{MCFOST} to compute the SED of HD\,121617 with the aim of constraining the following dust parameters: the dust mass M$_\mathrm{dust}$, the minimum grain size radius s$_\mathrm{min}$, and the power law of the grain size distribution $q$. In \texttt{MCFOST}, stellar parameters are defined by the the Kurucz model \citep{2003IAUS..210P.A20C} which corresponds the most to the SED. The selected model is the Kurucz model at T$_{\rm eff}=9500$\,K, log\,$g=5.0$ and the assuming the solar metallicity. The procedure to determined this model is presented in Appendix \ref{section_annex1}. The disk morphology is defined by the results of the MCMC analysis of Section \ref{section4} (though we assumed a disk with zero eccentricity) with $a=78.6$\,au, $i=43.4^{\circ}$, $\phi=240.0^{\circ}$,
$\alpha_\mathrm{out}=-5.8$, and $\alpha_\mathrm{in}=18.9$. For the sake of simplicity and given the small eccentricity ($\sim0.03$) we fix the eccentricity to zero. For the grain properties, we used the Mie theory (i.e., compact spherical grains), with a maximum grain size radius s$_\mathrm{max}=1000\,\mum$ and we used the optical constant of astrosilicate grains \citep{1984ApJ...285...89D}. Several studies of HR\,4796 shows that Mie theory is not always adapted to described the dust grains geometry \citep{2015ApJ...799..182P,2019A&A...626A..54M}. However, the measurement of the scattering phase function of a total intensity observation in required to constrain the dust grains geometry, in addition of the scattering phase function in polarimetry. Unfortunately the total intensity is not available for HD\,121617, yet.

Then \texttt{MCFOST} is used to produce a synthetic SED of the system with a log-spaced sampling of 300 wavelengths from $0.1\,\mum$ to $1500\,\mum$. This synthetic SED is finally converted into synthetic photometric points for specific filters for comparison with photometric data. 
We proceeded using \texttt{MCFOST} in the non-LTE configuration.
Since the disk should be optically thin, the infrared flux scales linearly with the total dust mass. Therefore, we fixed the dust mass and afterwards find the scaling factor $S_\mathrm{flux}$ that minimizes the residuals. A posteriori, we checked that the best fit solution indeed remains in the optically thin regime at all wavelengths with $\tau$ = 0.19 at $1,245\,\mum$ in the mid plane. The scaling factor $S_\mathrm{flux}$ is computed as:
\begin{equation}
     S_{flux} = \frac{  \sum\limits_{\lambda} \frac{F_{disk}(\lambda)}{\sigma(\lambda)^2} [ F_{obs}(\lambda) - F_{star}(\lambda) ]   }{ \sum\limits_{\lambda} \frac{F_{disk}(\lambda)^2}{\sigma(\lambda)^2} }, 
\end{equation}
where $F_\mathrm{obs}(\lambda)$ is the observed flux for the $\lambda$ filter, $\sigma(\lambda)$ the observed flux error for the $\lambda$ filter. $F_\mathrm{disk}(\lambda)$ and $F_\mathrm{star}(\lambda)$ are, respectively, the synthetic flux of the disk component and the star component for the $\lambda$ filter from \texttt{MCFOST}.

To determine the best values of s$_\mathrm{min}$ and $q$, they are sampled into a grid of values (Tab. \ref{mcfostparam}). 
The $\chi^2$ is computed with the infrared photometry in the same way as for the star parameters analysis in section \ref{section2}, including the following filters: WISE\,$W3$, WISE\,$W4$, PACS\,100, PACS\,160 and the ALMA observation at 1.33\,mm. We did not consider filters below $10\,\mum$ because the disk contribution at those wavelengths is negligible compared to that of the star.

\subsection{Results}
\label{section5.3}

\begin{figure*}[] 
\includegraphics[width=0.32\textwidth,clip]{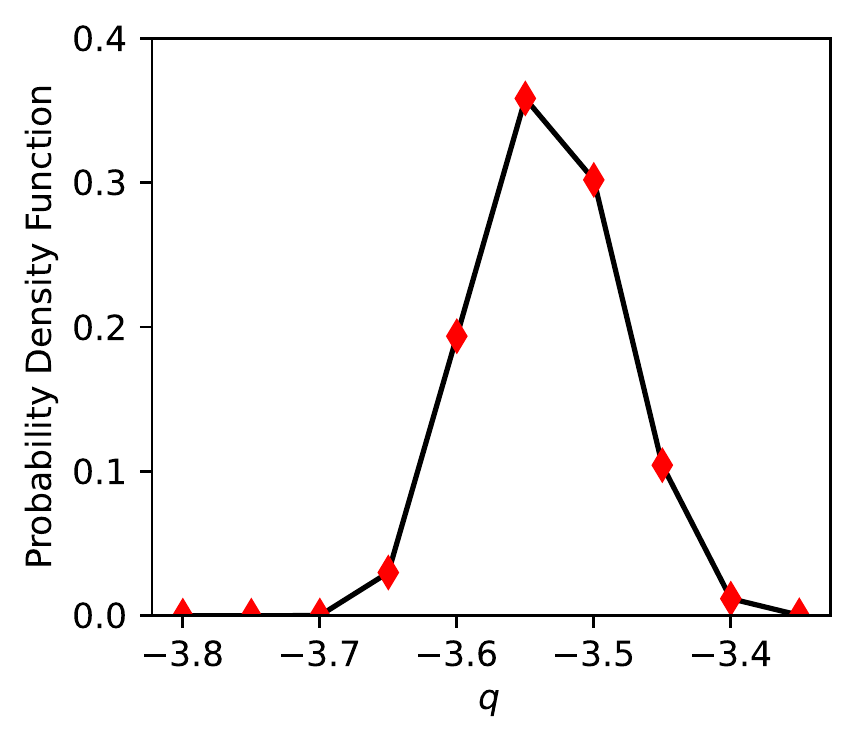}
\includegraphics[width=0.32\textwidth,clip]{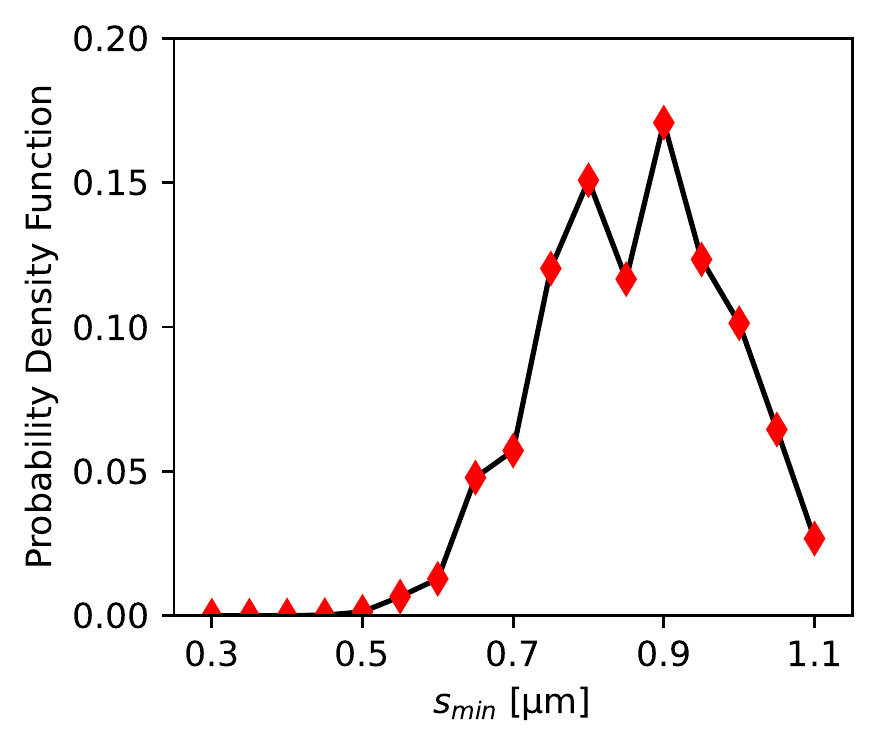}
\includegraphics[width=0.32\textwidth,clip]{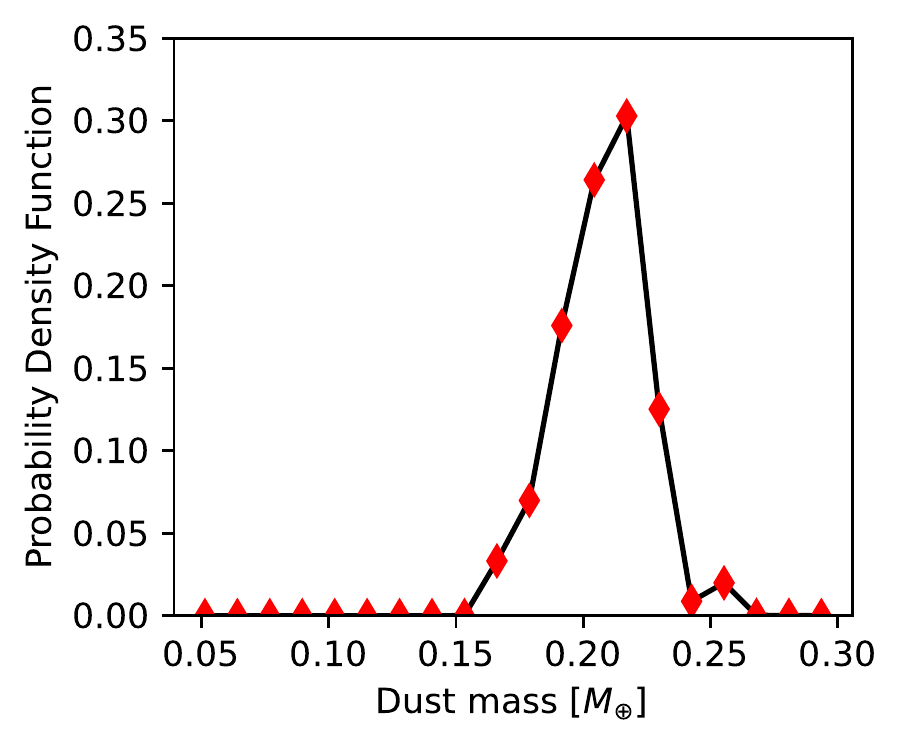}
\normalsize
\caption{Probability Density Function (PDF) for $q$, $s_\mathrm{min}$ and M$_\mathrm{dust}$.}
\label{PDF_MCFOST}
\end{figure*}

Table \ref{mcfostparam} shows the results of the analysis with \texttt{MCFOST}, with values for the best-fit model derived from the PDFs shown in Fig. \ref{PDF_MCFOST}. The central value is taken to be when the cumulated PDF reaches 50\% of the maximum while the error bars are for 16\% and 84\%, respectively. We find $s_\mathrm{min}\,=\,0.87^{+0.12}_{-0.13}\,\mum$, $q\,=\,-3.53$\,$\pm$\,$0.05$ and $M_\mathrm{dust}\,=\,0.21\pm0.02\,M_{\oplus}$.
As the dust mass is fixed via a scaling to the SED, we cannot directly compute the PDF for the mass. Instead, we randomly sample the output distribution of $M_\mathrm{dust}$ to produce a homogeneous grid of values, suitable to produce a PDF. This operation is repeated 1000 times for averaging purposes. 
These results are discussed in the next section.

\begin{table}[ht!]
\caption{Parameters and results of the \texttt{MCFOST} analysis.}
\footnotesize
\centering
\begin{tabular}{ l c c c}
\hline 
\vspace{0.0cm}
Parameter & Interval & Step & Best-fit model \\
\hline
$s_\mathrm{min}$ [$\mum$] & [0.30, 1.10] & 0.05 & $0.87^{+0.12}_{-0.13}$\\
\vspace{0.05cm}
 $q$      &  [-3.8, -3.35] & 0.05  & $-3.53$\,$\pm$\,$0.05$\\
\vspace{0.05cm}
M$_\mathrm{dust}$ [M$_{\oplus}$] & [0.27] & -- $^a$ & $0.21$\,$\pm$\,$0.02$\\
\hline
\end{tabular}
\flushleft
\vspace{0.1cm}
\textbf{Notes.} $^{\rm a}$ The dust mass M$_{dust}$ has only an initial mass and does not have a step value due to the method to compute it.
\normalsize
\label{mcfostparam}
\end{table}

\begin{figure}[] 
\includegraphics[width=0.49\textwidth,clip]{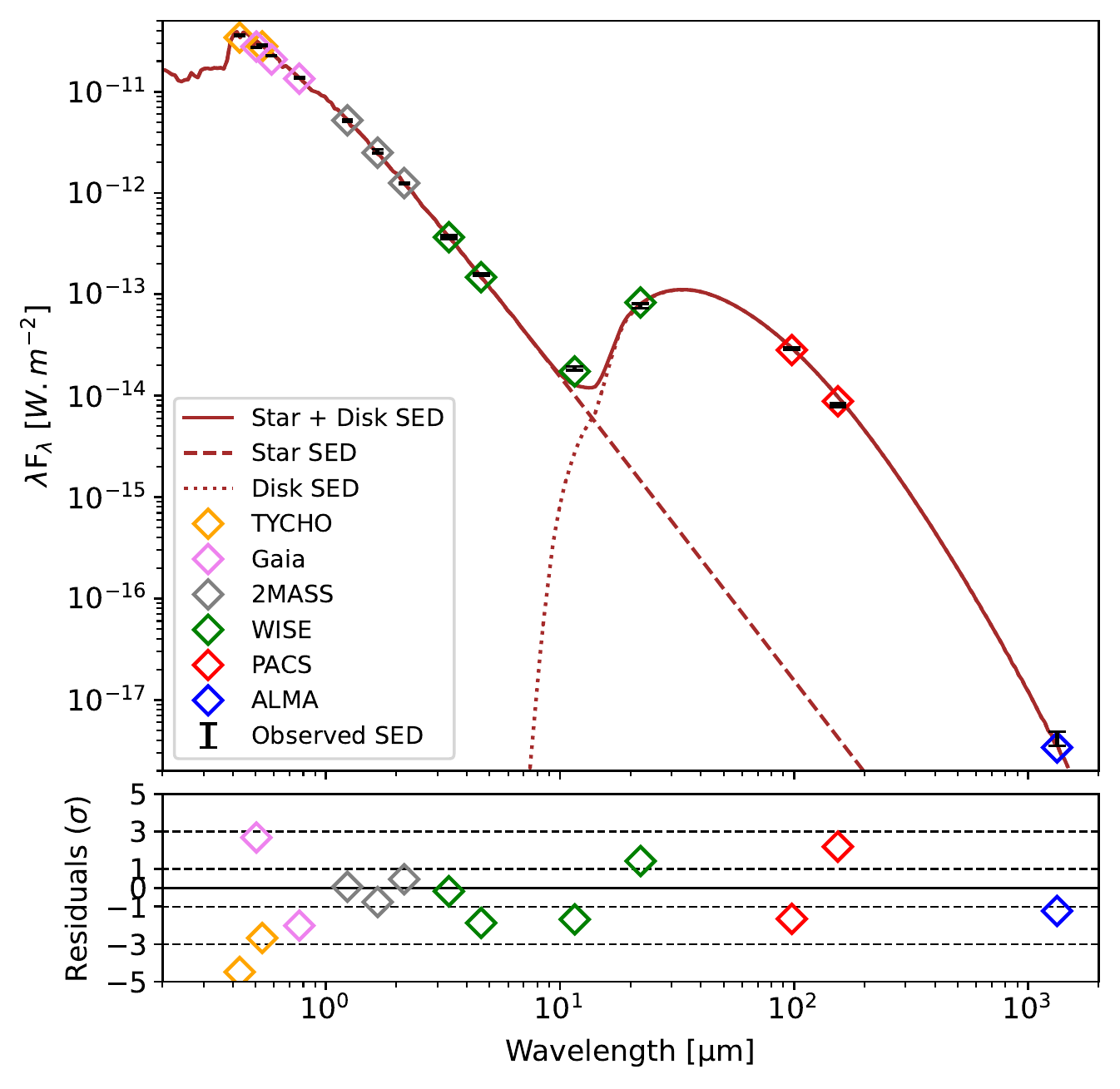}\\
\normalsize
\caption{SED of the HD 121617 system (top) and the residuals after subtraction of the best-fit model (bottom). The best model of the star is represented by the brown dashed line and that of the disk by the brown dotted line, while their sum is the brown solid line. The photometric points are represented with vertical black symbols including error bars and the instrument used for the observations are shown with diamonds of different colors.}
\label{bestfitsed}
\end{figure}

\section{Discussion}
\label{section6}

\subsection{Radial profile}\label{rad}

Our results lead to a high value of $\alpha_\mathrm{in}=18.9$, corresponding to a very sharp inner edge of the disk. Such sharp inner edges have been witnessed around several other systems \citep[see Table E.1 in][]{Adam2021}, and would indicate that "something" is shaping them, the most likely explanation being that a sharp inner edge corresponds to the outer limit of the chaotic region surrounding a yet-undetected planet just located inward of the belt \citep[e.g.][]{lagrange2012}. Note that, contrary to outer edges, radiation pressure or stellar wind do not place small grains in the dynamically "forbidden" region inward of the inner edge, which probably explains why sharp inner edges are more commonly observed than sharp outer ones \citep{Adam2021}. Poynting-Robertson drag could make small grains spiral inward, which could smooth out the sharpness of an inner disk edge, but the fraction of grains that can escape the main belt this way should be negligible for a very bright and collisionally-active disk such as HD 121617 \citep{Wyatt2005}.

For the outer edge of the disk radial profile, we get $\alpha_\mathrm{out}=-5.8$, corresponding to a slope for the radial surface density profile in $-4.8$. This is more than the canonical slope in $-1.5$ (or in $-3.5$ for the flux) that is expected in the outer regions beyond a collision-dominated belt of parent bodies, under the competing effect of collisional activity within the belt and radiation pressure (or stellar wind) placing small grains on highly eccentric orbits outside of it \citep{Strubbe2006,2008A&A...481..713T}.  Note, however, that this $-1.5$ slope is not supposed to be reached immediately beyond the main belt, but after a transition region, of relative width $\Delta r/r_0\sim0.2-0.3$, where the radial profile can be significantly steeper \citep{thebault2012b,thebault2014}. In the present case, the radial profile is constrained out to $1.1$\arcsec, after which the noise begins to dominate the signal (Fig.~\ref{radialprofile}), which is approximately $40\%$ outside of the peak radial location. This is slightly more than the theoretical width of the "natural" transition region beyond a collisional belt and could thus be interpreted as the signature of some additional removal process, such as dynamical perturbations by a planet or a stellar companion \citep{Thebault2012a, lagrange2012}. Another possible explanation for a sharp density drop beyond a belt of parent bodies could be that the belt is dynamically "cold", with a low collisional activity that creates a dearth of small grains with respect to larger ones \citep{2008A&A...481..713T}. However, this scenario appears unlikely in the present case, given the very high fractional luminosity ($\sim4.8\times10^{-3}$) of this disk, which should indicate a high level of dustiness. 

At any rate, our understanding of both the inner and outer edges would greatly benefit from future observations constraining the disk's radial profile further away from the main belt and also the potential presence of planets just inside or outside of it (e.g. with the Webb Space Telescope).

\subsection{Dust mass}

In debris disks the bulk of the mass is carried by large bodies that we cannot see neither with the full SED nor in the submm. However, we have access to the dust mass via mm-observations.

The value of $M_\mathrm{dust}=0.21\,M_{\oplus}$ we find is slightly higher than the dust mass obtained by \citet{2017ApJ...849..123M} from the flux density at 1.3mm, $0.12\,M_{\oplus}$\footnote{We apply the correction of the distance to the initial value obtained by \citet{2017ApJ...849..123M}, $1.4\,\times\,10^{-1}\,M_{\oplus}$.}. The difference between the two values is likely to be due to the fact that we use the whole SED to compute the mass whereas it is only based on the $1.3$\,mm flux in the other study. Moreover, we use slightly different assumptions for the dust composition and opacities, which can lead to this kind of differences.
However, these differences are relatively limited and our $M_\mathrm{dust}$ value remains compatible with previous estimates.
\\

\subsection{Size distribution and smallest grains}

The best-fit value found for $q$, the power law of the grain size distribution, is $-3.53$, which is remarkably close to the canonical value $-3.5$ expected for infinite self-similar collisional cascades at steady-state \citep{Dohnanyi1969}, and is fully compatible with size distribution profiles found in more realistic numerical explorations of collisional debris disks \citep{2007A&A...472..169T,2013A&A...558A.121K}\footnote{Given the simplicity of the present analysis we chose to ignore more complex features found in realistic size distributions such as wavy patterns}.

As for the minimum grain size $s_\mathrm{min}=0.87\,\mum$ that we find, it is useful to compare it to the blowout size $s_\mathrm{blow}$ corresponding to the grain size for which the ratio $\beta$ between the radiation pressure ($F_\mathrm{rad}$) and the gravitation force ($F_\mathrm{grav}$) is equal to 0.5. Grains smaller than $s_\mathrm{blow}$ that are produced from parent bodies on circular orbits are ejected out of the system due to radiation pressure. $\beta$ is defined by: 
\begin{equation}
    \beta = 0.5738 Q_\mathrm{pr} \left( \frac{1\,{\rm g \,cm}^{-3}}{\rho_\mathrm{d}} \right) \left( \frac{1\,\mum}{s} \right) \frac{L/L_{\odot}}{M/M_{\odot}},
    \label{eqkrivov}
\end{equation}
where $L$ is the stellar luminosity, $G$ the gravitational constant, $M$ the stellar mass, $c$ the speed of light, $Q_\mathrm{pr}$ the radiation pressure efficiency averaged over the stellar spectrum, and $\rho_\mathrm{d}$ the dust density. We assume $Q_\mathrm{pr}=1$ for geometric optics approximation and $\rho_\mathrm{d}=3.3\,{\rm g.cm}^{-3}$ for typical astrosilicate grains \citep{2009A&A...507.1503K}. With the stellar parameters from Appendix \ref{section_annex1}, we obtain $s_\mathrm{blow}=2.91\,\mum$, which is $\sim 3$ times larger than $s_\mathrm{min}$. 

We cannot rule out that this difference is due to assumptions made about the grain geometry, such as $Q_\mathrm{pr}=1$, or chemical composition (pure astrosilicates). \citet{2015MNRAS.454.3207P} show indeed that those parameters can have an important contribution to the blowout size. In particular, for astrosilicate the value of $Q_\mathrm{pr}$ varies between 2 and 0.2 as function of the wavelength \citep[Fig. 2]{2015MNRAS.454.3207P}. However, other dust compositions (such as carbon, ice or a mix) have lower dust densities, implying a higher blowout size. 

If, however, this discrepancy between $s_\mathrm{blow}$ and $s_\mathrm{min}$ is real, then HD\,121617 would join the handful of systems, such as HD\,32297, AU\,Mic, HD\,15115 or HD\,61005 \citep{thebault2019}, for which a significant presence of grains with $\beta>0.5$ has been inferred. Such a presence has often been interpreted as due to violent and/or transient events, such as the catastrophic breakup of a large planetesimal \citep{johnson2012,kral2015} or a so-called collisional avalanche \citep{grigorieva2007,thebault2018}, with the caveat that such events might be short lived and statistically unlikely. However, the numerical exploration of \citet{thebault2019} has shown that, for bright disks around A stars, there is a significant population of $s<s_\mathrm{blow}$ grains even for a "standard" debris disk at collisional steady-state. HD\,121617, with a $f_\mathrm{d}=4.8\times10^{-3}$ disk around an A1V central star, would nicely fit into this category.

Another possible explanation for the presence of small $\beta>0.5$ particles could be the effect of the gas that has been unambiguously detected in this system. Gas drag could indeed slow down the outward motion of unbound grains \citep{2019A&A...630A..85B} and push small micron-sized bound grains in regions where collisional activity is lower and lifetimes are longer \citep{2022arXiv220208313O}. This potential effect of gas on the observed dust, which could also affect the profile of the belt's inner and outer edges, is discussed in more detail in the next subsection.

\subsection{Effect of gas on the dust grains}

The results we have obtained so far do not necessarily require the presence of gas to be explained, but we note that large amounts of CO ($\sim 2 \times 10^{-2}$ M$_\oplus$) are present in the system \citep{2017ApJ...849..123M}, which may lead to the gas dragging the smallest dust grains \citep{2001ApJ...557..990T}. To verify this, we calculate the stopping time of grains of size $s$, of bulk density $\rho_\mathrm{d}$, located in a disk at 78\,au of radial extent $\Delta R$, bathed in a total mass of gas $M_{\rm gas}$, which leads to

\begin{equation}
\label{stopeq}
T_\mathrm{s} \sim 6 \, \left( \frac{\rho_\mathrm{d}}{2 \, {\rm g/cm}^3} \right) \left( \frac{s}{1 \, \mum} \right)\left( \frac{M_{\rm gas}}{2 \times 10^{-2}\,{\rm M}_\oplus} \right)^{-1} \left( \frac{\Delta R}{50\,{\rm au}} \right).
\end{equation}

\noindent A particle with a stopping time close to 1 orbital period will react to gas in about one dynamical timescale (the dust grain is said to be marginally coupled). Therefore, we find that the smallest grains in the system ($\sim$ 1 $\mum$) can begin to respond to the presence of gas within tens of orbital timescales considering only CO is present. In the presence of radiation pressure, these grains are expected to move slowly outward \citep[e.g.][]{2001ApJ...557..990T}. The final parking location of these grains is where $\beta=\eta$, where $\eta$ is related to the gas pressure gradient and sets the gas velocity to $v_\mathrm{k} \sqrt{1-\eta}$ ($v_\mathrm{k}$ being the Keplerian velocity). However, they can be destroyed by collisions faster than they reach this parking location, depending on the mass of gas and the density of the dust disk \citep{2022MNRAS.513..713O}. Overall, the surface brightness radial slope is expected to be shallower than usual (i.e. closer to -2 than -3.5) beyond the disk, in the halo. Before reaching this halo-like slope, there is an abrupt transition where the planetesimal belt stops, leading to surface brightness slopes $<-5$ over short radial distances. However, the latter is true even in the case where no gas is present \citep{2007A&A...472..169T}. Although the abrupt transition can be observed with SPHERE, the halo is diluted in the noise and its radial slope cannot be calculated from current observations. The main conclusion is that, even though gas is expected to be able to drag the smallest grains, its effects are not detectable on the radial profiles given current observations.  

It is expected that CO will photodissociate and create some carbon and oxygen. Depending on the amount of shielding and the viscosity of the gas disk, the number densities of carbon and oxygen may exceed that of CO \citep[e.g.][]{2019MNRAS.489.3670K}. In the case where there is much more mass than $\sim 2 \times 10^{-2}$ M$_\oplus$ (e.g. carbon and oxygen dominate), we still expect the same general conclusion on the radial profile (even if H$_2$ dominates, i.e. primordial origin). The main difference will be that small grains will move outward more quickly and have a better chance of reaching their parking place before being destroyed by collisions. 

However, one major difference is in the vertical profile of the grains. In the case of low gas masses where CO dominates the gas mass, vertical settling should only be significant for the smallest grains observed in scattered light, but it is difficult to spot in such an inclined system \citep{2022MNRAS.513..713O}. For more massive gas disks, the larger grains will have time to settle, creating needle-like disks in the sub-mm as well \citep{2022MNRAS.513..713O}. This difference would be easier to spot for an edge-on system but is difficult given the geometry of the disk around HD\,121617.

Finally, one may wonder whether the steep inner edge may be explained by the presence of gas. As explained in section \ref{rad}, many disks observed in scattered light have steep inner edges without gas being detected, and other causes such as the presence of a planet (see section \ref{rad}) provide a good explanation for their origins. However, we note that the gas helps to carve this steep inner edge as the smallest micron-sized grains observed in scattered light would be pushed outward on $\sim$ 10 dynamical timescales (see Eq.~\ref{stopeq}), which is to be compared to the timescale for refilling them, a collisional timescale. For the smallest grains, the collision timescale can be approximated by $(\tau \Omega)^{-1}$, which corresponds to 200 dynamical timescales, given the optical depth $\tau \sim 5 \times 10^{-3}$. The order of magnitude difference between the two timescales means that the smallest micron-sized grains would be slightly depleted in the innermost region, which cannot be refilled by grains coming from closer in regions. Finer observations and numerical simulations dealing with both dynamics and collisions in the presence of gas would be needed to draw more solid conclusions as to whether gas drag can really create steeper inner edges.


\section{Conclusion}
\label{section7}

Using the polarimetric mode of the VLT/SPHERE-IRDIS instrument, we have resolved the gas-rich debris disk around the star HD\,121617 for the first time in scattered light. The observations were made in 2018 using the dual-beam polarimetric imaging (DPI) mode in \textit{J} band with a corresponding angular resolution of 0.03\arcsec \, (or $\sim 3.5$\,au), an order of magnitude better than previous ALMA observations in the mm \citep{2017ApJ...849..123M}.

The high contrast of the images coupled with the high angular resolution allows us to significantly improve the previous constraints on the disk morphology. We fit the image using a disk located at a semi-major axis $a$, with an inclination $i$ and an eccentricity $e$ with a density made up of two power laws $\alpha_{\rm in}$ and $\alpha_{\rm out}$. Using the radiative transfer code \texttt{DDiT} in an MCMC fashion, we find that the best fit is $a=78.6\pm0.6$ au, $i=43.4\pm0.8^{\circ}$, $e=0.03\pm0.01$, $\alpha_{\rm in}=18.9\pm2.3$ and $\alpha_{\rm out}=-5.8\pm0.4$. We also constrain the position angle of the major axis to $\phi=240.0\pm0.9^{\circ}$, the anisotropic scattering coefficient to $g=0.60\pm0.04$ (if positive) and the argument of pericenter (if eccentric) to $\omega=131.0\pm15.1^{\circ}$.

Taking advantage of these new geometric constraints to mitigate degeneracies in the SED fitting process, we derive the dust properties by fitting the full SED of the system. We find that the minimum grain size is best fitted by $s_{\rm min}=0.87^{+0.12}_{-0.13} \, \mum$, a value that appears smaller than the blowout size. If real, it would not be surprising as the disk around HD 121617 has a very high fractional luminosity close to $5\times 10^{-3}$ and in this case an overabundance of small unbound grains of submicron size is expected \citep{thebault2019}. The presence of large amounts of gas as in this system may also be able to slow down these unbound grains which can then accumulate even more \citep[e.g.][]{2019A&A...630A..85B}. The SED fit also leads to a size distribution slope $q=-3.53\pm0.05$ which is naturally expected in a collision-dominated debris disk \citep[e.g.][]{2007A&A...472..169T,2013A&A...558A.121K}. Finally, we constrain the dust mass to $M_\mathrm{dust}=0.21\pm0.02\,M_{\oplus}$.

One of the main constraints on the disk morphology is that the density profile of the inner edge is radially very steep (power law in $\sim r^{20}$), which could be due to the presence of a yet unseen planet near the inner edge of the disk. We also note that, qualitatively, the presence of gas in this system could be the reason for the sharp inner edge of the dust profile. The outer edge is also sharper than predicted by steady-state models of debris disk halos, but our constraint should be taken with caution as it only concerns the very early parts of the halo (up to 1.1\arcsec), where a sharp transition region is naturally expected. Further observations targeting the halo at larger distances (e.g. JWST) would be needed to infer the presence of a planet or to see the effect of gas on the outer profile of the surface brightness at large distances, which is expected to be shallower than -3.5 and closer to -2 \citep{2022MNRAS.513..713O}.

\begin{acknowledgements}
C.\,P. acknowledges support by the French National Research Agency (ANR Tremplin-ERC: ANR-20-ERC9-0007 REVOLT).
C.\,P. also acknowledge financial support from Fondecyt (grant 3190691).
J.\,O. acknowledge financial support from Fondecyt (grant 1180395).
M.\,M. acknowledge financial support from Fondos de Investigaci\'on 2022 de la Universidad Vi\~na del Mar.
G.\,M.\,K. was supported by the Royal Society as a Royal Society University Research Fellow.
C.\,P., J.\,O., M.\,M., A.\,B. and D.\,I. acknowledge support by ANID, -- Millennium Science Initiative Program -- NCN19\_171.
This publication makes use of VOSA, developed under the Spanish Virtual Observatory (\url{https://svo.cab.inta-csic.es}) project funded by MCIN/AEI/10.13039/501100011033/ through grant PID2020-112949GB-I00. VOSA has been partially updated by using funding from the European Union's Horizon 2020 Research and Innovation Programme, under Grant Agreement nº 776403 (EXOPLANETS-A).
This research has made use of the SVO Filter Profile Service (http://svo2.cab.inta-csic.es/theory/fps/) supported from the Spanish MINECO through grant AYA2017-84089.
This work has made use of data from the European Space Agency (ESA) mission {\it Gaia} (\url{https://www.cosmos.esa.int/gaia}), processed by the {\it Gaia} Data Processing and Analysis Consortium (DPAC, \url{https://www.cosmos.esa.int/web/gaia/dpac/consortium}). Funding for the DPAC has been provided by national institutions, in particular the institutions participating in the {\it Gaia} Multilateral Agreement.
We thank the referee for they helpful comments.

\end{acknowledgements}

\bibliography{bib}

\clearpage
\appendix

\section{Stellar properties}

\label{section_annex1}

\begin{figure}[ht!] 
\includegraphics[width=0.49\textwidth,clip]{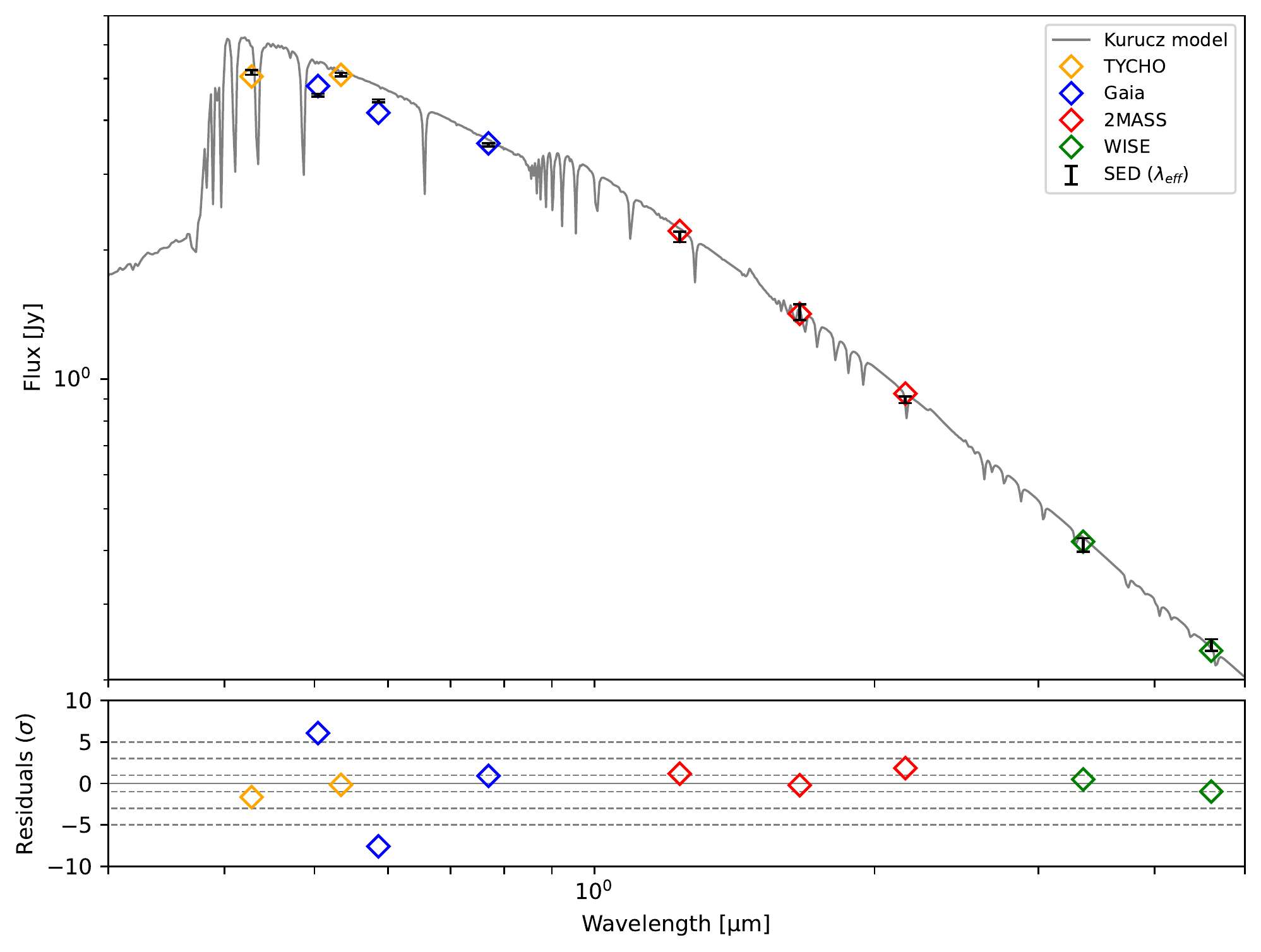}\\
\normalsize
\caption{Kurucz stellar spectrum for a star with $T_\mathrm{eff}$\,=\,9500\,K and log\,\textit{g}\,=5.0 (grey line), the synthetic SED compute with the Kuzucz model (colored diamonds) and the observational SED (black dots). The lower panel show residuals between synthetic photometries and the SED in number of $\sigma$. 
}
\label{SED_stellar_param}
\end{figure}

We determined the stellar parameters by comparing the observed SED, corrected to the extinction, to Kurucz models \citep{2003IAUS..210P.A20C}. We used a grid of stellar spectra with an effective temperature range between $8000$\,K and $9750$\,K, with a sampling of $250$\,K and a log\,$g$ ranging between $3$ and $5$ with a sampling of $0.5$.
For each Kurucz stellar spectrum, we computed the synthetic photometry for the $10$ filters used (getting their properties from the Spanish Virtual Observatory (SVO) filter profile service (\citealp{2012ivoa.rept.1015R,2020sea..confE.182R}). The filters used are TYCHO ($B$ and $G$), Gaia DR2 ($Gbp$, $G$ and $Grp$), 2MASS ($J$, $H$ and $Ks$) and WISE ($W1$ and $W2$) 
The synthetic SED is fitted to the observed SED where the scaling factor corresponds to the dilution factor (R$_\star/d_\star)^2$, where $d_\star$ is the known distance in parsec and R$_\star$ the stellar radius in units of solar radius. The best stellar spectrum is obtained by minimizing the goodness of fit for each spectrum of the grid.
The stellar parameters of the best fit solution are T$_{\star}$\,=\,$9500$\,K, log\,\textit{g}\,=\,$5.0$, R$_{\star}$\,=\,$1.54$\,R$_{\odot}$ and L$_{\star}$\,=\,$17.64$\,L$_{\odot}$. 
R$_{\star}$ is determined using the dilution factor. From the stellar luminosity, we also derived the stellar mass from the mass-luminosity relation for main sequence star L$_{\star}$\,=\,M$_{\star}^{3.5}$, which gives M$_{\star}$\,=\,$2.27$\,M$_{\odot}$.
The best Kurucz stellar spectrum is shown in Fig.\,\ref{SED_stellar_param} and used in the section \ref{section5} for the \texttt{MCFOST} analysis.

\section{MCMC analysis of the \texttt{DDiT} model}

\begin{figure}[] 
\includegraphics[width=0.49\textwidth,clip]{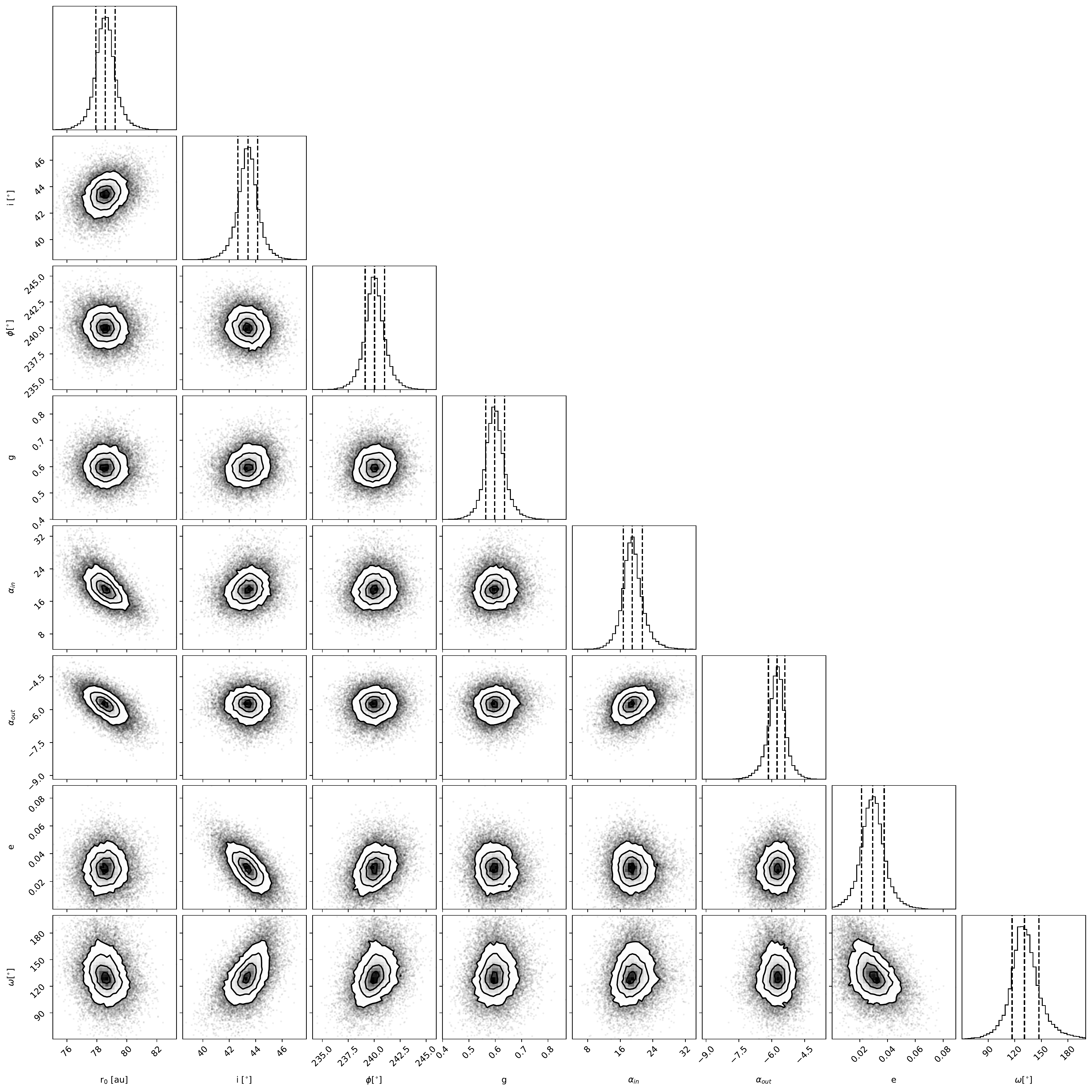}
\normalsize
\caption{Corner-plot for the MCMC analysis. From left to right and top to bot: $a$ ($r_0$), $i$, $\phi$, $g$, $\alpha_\mathrm{in}$, $\alpha_\mathrm{out}$, $e$ and $\omega$. Middle dash-lines are the more probable value and external dash-lines are the 1-$\sigma$ error.}
\label{cornerplot}
\end{figure}

\end{document}